\documentclass[
    prl,
    aps,
    twocolumn,
    amsmath,amssymb,
    longbibliography,
]{revtex4-2}

\usepackage[utf8]{inputenc}
\usepackage[english]{babel}
\usepackage[dvipsnames]{xcolor}
\usepackage{hyperref}
\usepackage{graphicx}
\usepackage{dcolumn} 
\usepackage{bm}

\hypersetup{
    colorlinks=true,
    linkcolor=blue,
    citecolor=blue,
    urlcolor=blue
}

\emergencystretch=\maxdimen
\makeatletter
\def\blfootnote{\xdef\@thefnmark{}\@footnotetext}
\makeatother

\begin{document}

\title{Threshold-free estimation of entropy from a Pearson matrix}

\author{H.~Felippe$^{1,2}$*\blfootnote{* Corresponding author:
        \href{mailto:h.felippe@fisica.ufrn.br}{h.felippe@fisica.ufrn.br}},
        A.~Viol$^{3}$,
        D.B.~de~Araujo$^{4}$,
        M.G.E.~da~Luz$^{5}$,
        F.~Palhano-Fontes$^{4}$,
        H.~Onias$^{4}$,
        E.P.~Raposo$^{6}$,
        G.M.~Viswanathan$^{1,7}$
}

\affiliation{
  $^{1}$ Department of Physics, 
         Federal University of Rio Grande do Norte - 
         Natal--RN, 59078-970, Brazil\\
  $^{2}$ Department of Network and Data Science, 
         Central European University - 
         Vienna, 1100, Austria\\
  $^{3}$ Cognitive Neuroscience,
         Scuola Internazionale Superiore di Studi Avanzati - 
         Trieste, 34136, Italy\\
  $^{4}$ Brain Institute, 
         Federal University of Rio Grande do Norte - 
         Natal--RN, 59076-550, Brazil\\
  $^{5}$ Departamento de F\'{i}sica,
         Universidade Federal do Paran\'{a} -
         Curitiba--PR, 81531-980, Brazil\\
  $^{6}$ Laborat\'{o}rio de F\'{i}sica Te\'{o}rica e Computacional,
         Departamento de F\'{i}sica, 
         Universidade Federal de Pernambuco -
         Recife--PE, 50670-901, Brazil\\
  $^{7}$ National Institute of Science and Technology of Complex Systems,
         Federal University of Rio Grande do Norte - 
         Natal--RN, 59078-970, Brazil
}

\date{\today}


\begin{abstract}
There is demand in diverse fields for a reliable method of estimating
the entropy associated with correlations. The estimation of a unique
entropy directly from the Pearson correlation matrix has remained an
open problem for more than half a century. All existing approaches
lack generality insofar as they require thresholding choices that
arbitrarily remove possibly important information. Here we propose an
objective procedure for directly estimating a unique entropy of a
general Pearson matrix. We show that upon rescaling the Pearson matrix
satisfies all necessary conditions for an analog of the von Neumann
entropy to be well defined. No thresholding is required. We
demonstrate the method by estimating the entropy from neuroimaging
time series of the human brain under the influence of a psychedelic.
\end{abstract}

\maketitle

An important open problem in many fields concerns how to estimate the
entropy from the Pearson correlation matrix of an arbitrary data
set~\cite{alb02,ana09,man16}.  This problem has wide and immediate
applicability to diverse phenomena, \textit{e.g.}, correlated asset
prices~\cite{alm19,cha21}, diffusion processes~\cite{gom08}, and brain
networks~\cite{bul09,san19,nic20}.  The Shannon entropy solves the
simpler problem of how to calculate the entropy given a probability
distribution~\cite{shannon}. However, attempts to extend this formalism
to estimate the entropy of a correlation matrix have required admittedly
ad hoc methods, such as the thresholding of the correlation matrix to
generate a new symmetric and unweighted adjacency matrix consisting only
of 0's and 1's~\cite{lat01,yan18}. The resulting adjacency matrix
represents a simple graph, which can then be studied using topological
metrics. These methods have led to significant advances in diverse
fields, examples of which include applications to market volatility in
financial crisis~\cite{kum12}, and to scale-free and small-world brain
networks~\cite{egu05,bas06,gal12}.  Nevertheless, they have drawbacks
that motivate the search for better approaches~\cite{rub11}. 

The main limitation of the existing methods is that the thresholding
operation inevitably leads to loss of information~\cite{kuk20}.
Consider, for example, the set of all possible $2\times 2$ Pearson
correlation matrices. Since a correlation coefficient can assume any
real value in the closed interval $[-1,1]$, there are an infinite number
of possible $2\times 2$ Pearson matrices.  In contrast, there are only
two corresponding adjacency matrices of size $2\times 2$ because a
simple graph with two nodes is either connected or disconnected.
Significant information can thus be lost due to thresholding. As an
important example, we mention that in neuroimaging data the results are
not always robust with respect to the choice of
thresholding~\cite{lan13,gar15}.  Insofar as unweighted matrices are
used, these methods are only indirectly measuring the entropy of
correlations, generally via topological metrics such as network
motifs~\cite{tag14}, degree distribution~\cite{vio17}, and geodesic
diversity~\cite{vio19}.  There is thus a growing and vexing demand for a
better method of estimating entropy.

\begin{figure*}
\includegraphics[scale=1.15]{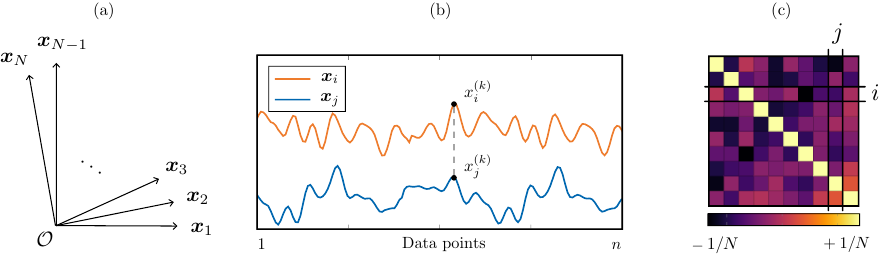}
\caption{
    The construction of a density matrix from a Pearson correlation
    matrix.  
    (a) The system of interest is partitioned into $N$ interacting
    components or nodes.
    (b) The correlation between nodes $i$ and $j$ is then calculated.
    For each $i$ node there are $n$ data points $x_i^{( k)}$ where
    $k=1,\dots,n$.  The correlation $R_{ij}$ between nodes $i$ and $j$
    is given by Eq.~(\ref{eq:rij}), and lies in the interval $-1\leq R_
    {ij}\leq 1$. 
    (c) For each correlation coefficient $R_{ij}$, we calculate
    $\rho_{ij}=R_{ij}/N$, yielding the matrix $\bm{\rho} =\bm{R}/N$.
    This matrix $\bm\rho$ fully satisfies the properties of a density
    operator, \textit{i.e.}, (i) hermiticity, (ii) unit trace, and (iii)
    positive semidefiniteness. Finally, we calculate the entropy from
    this density matrix via Eq.~(\ref{eq:trace}).
}
\label{fig:brain}
\end{figure*}

Here, we demonstrate that the problem of how to calculate the entropy of
a correlation matrix can be solved without the need of an adjacency
matrix or ad hoc schemes. Let $\bm{R}$ be a general $N\times N$ Pearson
correlation matrix and let $\bm{\rho}=\bm{R} /N$.  We show below that
$\bm{\rho}$ has $N$ non-negative eigenvalues $\lambda_j$ that add up to
$1$, so that $\bm{\rho}$ has unit trace  (\textit{i.e.},
$\mathrm{tr}\bm{\rho} =1$). If we interpret these eigenvalues as a
probability distribution, then we can define the Shannon entropy for
this distribution according to
\begin{equation}
	S(\bm{\rho}) = -\sum_{j=1}^N \lambda_j \log\lambda_j~.
	\label{eq:shannon}
\end{equation}
\noindent
In terms of the trace of a matrix, we can equivalently write the above
as 
\begin{equation}
	S(\bm{\rho}) = -\mathrm{tr}(\bm{\rho}\log\bm{\rho})~.
	\label{eq:trace}
\end{equation}

The reader familiar with quantum statistical mechanics will immediately
recognize Eq.~(\ref{eq:trace}) as an analogue of the von Neumann
entropy for a density operator $\bm{\rho}$ for  an ensemble of states
with probabilities $\lambda_j$~\cite{ohy93}.  Indeed, the problem of
calculating the entropy of a Hermitian matrix with non-negative
eigenvalues and with unit trace was solved almost a century ago by John
von Neumann~\cite{von27} and Lev Landau~\cite{lan65}. However, a form of
entropy defined in terms of a density operator had not been applied to
Pearson correlation matrices until now.  This procedure will thereby
allow the calculation of the entropy from the Pearson matrix (see
Fig.~\ref{fig:brain}).

\medskip

\textbf{Direct estimation of entropy from the Pearson matrix.~--}
We first review the definition of the Pearson correlation matrix.
Consider $N$ data sets, each with a population of $n$ measurements. Let
\mbox{$\bm{x} _i = (x_i^{(1)},\dots,x_i^{(n)})$} denote the sequence of
$n$ measurements of the $i$-th data set. Let $\langle ~ \cdot ~ \rangle$
denote the average taken over the population of $n$ points, so that
$\bm{X}_i = \bm{x}_i - \langle \bm{x}_i \rangle$ represents  deviations
from the mean $\langle \bm{ x}_i \rangle$ and $\sigma_i^2=\langle
\bm{X}_i^2 \rangle$ is  the variance of $\bm{x}_i$.  The correlation
between the $i$-th and $j$-th data sets is then given by the Pearson
correlation coefficient
\begin{equation}
  R_{ij}
  = \frac1 {n}~  \sum_{k=1}^n
  \Bigg[
    \frac
    {{x}_i^{(k)} -
  \langle \bm{x}_i \rangle }
{\sigma_i }
\Bigg]
  \Bigg[
  \frac
    {{x}_j^{(k)} -
  \langle \bm{x}_j \rangle}
{\sigma_j}
\Bigg]
  = \frac{ \langle \bm{X}_i \bm{X}_j \rangle}{ \sigma_i
          \sigma_j }
        ~. \label{eq:rij}
\end{equation}

\medskip\noindent
The Pearson correlation matrix is the $N\times N$ matrix $\bm{R}$ with
entries $R_{ij}$ as in Eq.~(\ref{eq:rij}).

Given a Pearson correlation matrix $\bm{R}$,  the entropy is often
estimated by using a different (but related) object as a proxy, viz., a
binary adjacency matrix $\bm{A}$. The latter is defined by imposing a
threshold value $\xi $ over the entries of $\bm{R}$. Specifically, one
sets \mbox{$A_{ij} = 1$} whenever $|R_{ij}| \geq \xi $ for   $i\neq j$,
and \mbox{$A_{ij} = 0$} otherwise~\cite{liu08}. The matrix $\bm{A}$
defines a graph with $N$ nodes and undirected and unweighted edges. The
entropy estimation, then, becomes a matter of computing the Shannon
entropy of certain probability distributions embedded in the graph
structure of $\bm{A}$, as explained earlier. We note that, in network
analysis, the entropy in Eq.~(\ref{eq:trace}) has previously been used
with adjacency matrices~\cite{passerini-2009,estrada-2014} and networks
in general~\cite{man13}, rendering a properly sub-additive spectral
entropy for model selection of complex
networks~\cite{man16,man15,ghavasieh}.  It has also been analyzed in the
context of measures of fit of multiple latent
variables~\cite{golino2021entropy}.  However, it has never been applied
directly to Pearson correlation matrices.

Numerous properties of correlation matrices can be extracted from their
thresholded counterparts, partly due to the information that remains
stored in the topology of simple graphs~\cite{vio21}. Hence, these
methods have become pivotal and ubiquitous in the study of complex
systems~\cite{kum12,egu05,bas06}.  
In neuroscience, for instance, different thresholding schemes are
extensively used to deduce the structural and functional connectivity of
the brain, despite the selected method possibly influencing over 
and misleading the
neuroimaging data analysis~\cite{bernadette,drakesmith} (for context,
see a thorough review of thresholding in network
neuroscience~\cite{hallquist,fornito}).
But there are yet other drawbacks to
this approach, in addition to the limitations previously
mentioned~\cite{rub11,kuk20,lan13,gar15}.  Not only can thresholding
lead to loss of information, it can even inject spurious complexity into
genuinely random networks~\cite{can20}. For these reasons, such methods
require caution when applied  to real-world networks~\cite{nic20}.

In order to overcome these difficulties, we develop a way to estimate
the entropy directly from the Pearson correlation matrix, without
relying on an adjacency matrix. Our method rests on the following
mathematical foundation: if $\bm{R}$ is an arbitrary $N\times N$ Pearson
correlation matrix, then
\begin{equation}
  \label{eq:eq-R-q03475983475}
  \bm{\rho}={\bm{R} \over N}
\end{equation}
satisfies all the conditions for a density operator, namely,
(i)~hermiticity, (ii)~unit trace, and (iii)~positive semidefiniteness.
To see this, observe that $\bm{R}$ has real and symmetric entries, from
which follows claim~(i).  Next, the diagonal entries are given by
\begin{equation}
	R_{ii}	= \frac {\langle \bm{X}_i^2 \rangle}{\sigma_i^2}
			= 1~,
\end{equation}
\noindent
so that $\mathrm{tr} \bm{R} = N$, leading to claim~(ii). For~(iii), it
suffices to show that $\bm{R}$ is positive semidefinite. In other words,
we must show that $\bm{v}  \cdot \bm{R} \, \bm{v} \geq 0$ for all
$\bm{v} \in \mathbb{R}^N$.  In fact, the positive semidefiniteness of
Pearson correlation matrices is well known.  Observe that
\begin{equation}
	\sum_{ij}^N v_i \left(\sum_k^n X_i^{(k)} X_j^{(k)}\right) v_j =
	\sum_k^n \left| \sum_i^N v_i X_i^{(k)} \right|^2 \geq 0 ~.
\end{equation}

We can now state our main result.  Given an arbitrary Pearson
correlation matrix, the entropy can be calculated from
Eq.~(\ref{eq:trace}), with $ \bm \rho$ given by
Eq.~(\ref{eq:eq-R-q03475983475}).  We remark in passing that many
other correlation matrices (\textit{e.g.}, Spearman, Kendall) will
satisfy analogous conditions, so long as they consist of real
symmetric entries~\cite{bou12}.  For normalized time series this
definition of the density operator is the same as the covariance
matrix divided by its trace.  Furthermore, one can readily verify that
$ S(\bm{\rho})=\log N$ is the maximal entropy value for a system of
$N$ uncorrelated data sets.

In conceptual terms, the method here proposed is a mathematical device
that yields a set of scaled eigenvalues, in such a way that makes it
possible to assign a well defined entropy to any given Pearson
correlation matrix.  This entropy is the analog of the von Neumann
entropy for a density operator $\bm \rho$ and is a continuous
non-negative real functional of the Pearson matrix.  However, apart
from the special case where $\bm\rho$ would correspond to a pure state
in quantum mechanics, the matrix \mbox{$\bm{\rho} = \bm{R}/N$}
corresponds to a mixed state in quantum statistical mechanics, such
that it encodes classical probabilities that are precisely the
eigenvalues of the scaled Pearson correlation matrix.

Since probability distributions (and density operators) become
coarse-grained (hence effectively renormalized) in the presence of
noise, the entropy of two different data sets with similar levels of
added noise can still be usefully compared. Therefore, this is an
advantage of our method given that it is robust to the addition of
noise.  Moreover, although the Pearson coefficient is a pair
correlation coefficient, it is still sensitive to nonlinear effects
because it is the expected value of a bilinear quantity.

It is worth pointing out a crucial difference between the method
proposed here and the other thresholding-based methods of calculating
entropy.  Whereas the proposed entropy is a continuous function
(functional) of the Pearson matrix, the usual thresholding-based
entropies are discontinuous. To see this, recall that the Heaviside
step function is defined as
\begin{equation}
\Theta(x) = \left\{
  \begin{array}{l}
  0, \quad x\leq 0 \\
  1, \quad x>0 ~~~ .\end{array} \right.
\end{equation}
This function has a discontinuity at $x = 0$, so that any entropy
calculated via thresholding is necessarily a discontinuous function of
the Pearson matrix, due to the explicit or implicit use of the Heaviside
step function to perform thresholding.  In contrast, the proposed
entropy is a continuous functional of the Pearson matrix.

We thus have a robust and direct approach to obtain the entropy of
correlation matrices.  Crucially, at no point have  we introduced {\it a
priori} estimates, hence the resulting entropy is free from arbitrary or
extraneous assumptions (\textit{e.g.}, choice of threshold $\xi$). Of
course, one can additionally use thresholding to obtain a suitable
$\bm{A}$, but even then our protocol can still be useful. For example,
it can be used to obtain an exact expression for the entropy of an
idealized ranked set of correlated data.  Such results might then yield
a null hypothesis for subsequent statistical inference or undertake more
rigorous hypothesis testing.

In what follows, we apply our main result to illustrative examples
that are easy to understand, in order to further build intuition on
the subject.  We then apply our method to functional magnetic
resonance imaging (fMRI) time series of the human brain activity.  We
also have included an extra material where we discuss in some detail
the related topic of Guttman-scalable Pearson matrices for those
readers interested in further applications (see the 
\hyperlink{supp}{Supplementary Material}).

\medskip

\textbf{Examples of some Pearson matrices.~--}
Consider first the scenario {of $N$ series} where all $\bm{x}_i$ and
$\bm{x}_j$ are statistically independent, \textit{i.e.}, uncorrelated:
\mbox{ $\langle \bm{X}_i \bm{X}_j\rangle = \langle
\bm{X}_i^2\rangle\,\delta_{ij}$.} Thus from equation
Eq.~(\ref{eq:rij}), \mbox{$R_{ij} =\delta_{ij}$}, and, as expected, the
density operator admits the (completely mixed state) representation
$\bm{ \rho} = \bm{I}/N$, where $ \bm{I}$ is the $N \times N$ identity
matrix.  From Eq.~(\ref {eq:trace}) one immediately finds that $S=\log
N$. It is a simple exercise to show that this is the maximum possible
entropy for an $N\times N$ density matrix. 

Compare the above with the other extreme, {\it i.e.}, when all pairs are
perfectly autocorrelated, so that
\mbox{$\bm{X}_i/\sigma_i=\bm{X}_j/\sigma_j$} for all $i,~j$.  Then,
clearly, \mbox{$R_{ij}=1$}, so that $\bm R^2 = N \bm R$.  The matrix
$\bm{\rho}$ is thus idempotent: $\bm{\rho}^2=\bm{\rho}$. It follows
immediately that
\begin{equation}
    \mathrm{tr}(\bm{\rho} \log\bm{\rho}) =
            \mathrm{tr}(\bm{\rho} \log\bm{\rho}^2) =
                    2\mathrm{tr}(\bm{\rho} \log\bm{\rho})~,
\end{equation}
hence $\mathrm{tr}(\bm{\rho} \log\bm{\rho})=0$. Then we have that $S=0$,
which is the absolute minimum possible value (corresponding to a pure
state configuration).  Note that, in contrast to statistical moments,
the entropy is not strongly dependent on the tails of the distribution
of eigenvalues, since $\lambda_j \log \lambda_j \to 0 $ in the limit
$\lambda_j \to 0$.

Let us now focus on analytically solvable examples, in order to illustrate
better how the entropy depends on the entries of the Pearson matrix.
Consider thus the following situation which is an interpolation between
the cases of maximum and minimum entropies.

\begin{figure}[t]
\centering
\includegraphics[scale=0.85]{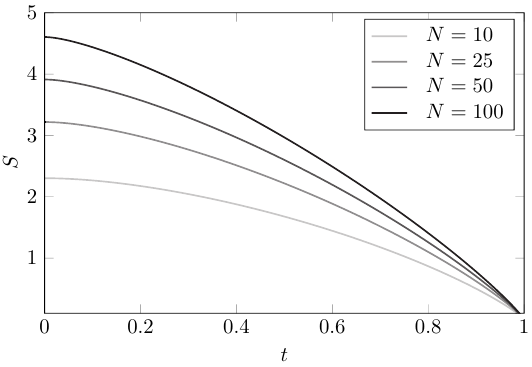}
\caption{
  Entropy $S$ as a function of the parameter $t$ for the example given
  by Eq.~(\ref{eq-eoirhwoihtgwt4h}).  The case $t=1$ represents maximal
  correlation ($S=0$) while $t=0$ represents uncorrelated series,
  leading to maximum entropy.  $S=\log N=\log 100\approx 4.6$, as
  expected. 
}
\label{fig:example1}
\end{figure}

Let the Pearson matrix now be given according to
\begin{equation}
    R_{ij}(t) = t  + \delta_{ij} (1-t)~.
\label{eq-eoirhwoihtgwt4h}
\end{equation}
This corresponds to a situation where any pair of distinct series have
autocorrelation $t\in [-1,1]$. It is easy to check that the density
operator $\bm{\rho}(t) = \bm{R}(t)/N$ for this Pearson matrix is well
behaved.  $N-1$ of the $N$ eigenvalues --- all but one of them --- are
identically equal to $(1-t)/N$. The last remaining eigenvalue equals
$(1+(N-1)t)/N$.  The entropy is thus
\begin{align}
  S(t) &= \log N  - {N-1\over N}(1-t)\log(1-t) \nonumber \\
  & \quad 
- {1 \over N} [1+(N-1)t] \log [1+(N-1)t] ~.
\end{align}
The special case $t=0$ recovers the maximum entropy $S=\log N$ while
$t=1$ recovers the minimum entropy $S=0$. Figure~\ref{fig:example1}
shows how the entropy varies with $t$ in between the two limiting
cases for $N=100$.

Our final example is the case of a modular Pearson matrix  $\bm{R_M}$
representing 2 weakly correlated sets of $N$ strongly correlated series.
Let the strong correlation be $t$ and let the weak correlation be
$\epsilon$.  Such a Pearson matrix is a $2N\times 2N$ matrix with a
$2\times 2$ block structure:
\begin{equation}
  \bm{R_M}(t,\epsilon) =
  \begin{bmatrix}
    \bm{R}(t) & \bm{\epsilon} \\
    \bm{\epsilon} & \bm{R}(t) 
  \end{bmatrix} ~~.
\label{eq-uheriuherb484848}
\end{equation}  
where the $\bm{R}(t)$ are given by Eq.~(\ref{eq-eoirhwoihtgwt4h}) and
the matrix $\bm{\epsilon}$ has every entry equal to $\epsilon$.  In this
case, all but 2 of the eigenvalues of the corresponding density matrix
$\bm{\rho}(t,\epsilon)$ are identically equal to $(1-t) /(2N)$. The
remaining 2 eigenvalues are $ (1+(N-1)t \pm N \epsilon) /(2N)$.  Hence,
the entropy is immediately obtained:
\begin{align}
  &S(t,\epsilon)=\nonumber
  \\ \nonumber
  & 
    \frac
1 {2 N} \bigg[
  - (\epsilon N+(N-1) t+1) \log
    \bigg(\frac{\epsilon N+(N-1) t+1}{2 N}\bigg)
    \\ 
    & \nonumber \quad \quad \quad 
    +(\epsilon N-N t+t-1)
    \log \left(\frac{-\epsilon N+(N-1) t+1}{2 N}\right)
    \\  
    &  \quad \quad \quad 
    +2 (N-1) (t-1)
    \log \left(\frac{1-t}{2 N}\right) \bigg] ~.   
\end{align}
Note that for the condition of positive semidefiniteness to be
satisfied, we require
$
    1+(N-1)t \pm N \epsilon \geq 0~,
$
so that for fixed $t$ we must have
\begin {equation}
    |\epsilon| \leq  {1+(N-1)t \over N  } ~.
\end {equation}

\begin{figure}[t]
\centering
\includegraphics[scale=0.85]{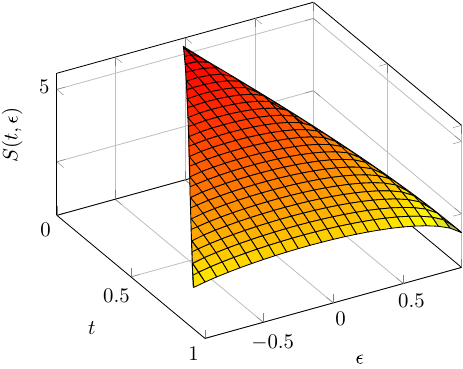}
\caption{
    Entropy $S$ as a function of the parameter $t$ and $\epsilon$ for
    the ``modular''  example given by Eq.~(\ref{eq-uheriuherb484848})
    for $N=100$. 
}
\label{fig:example2}
\end{figure}

Figure~\ref{fig:example2} shows the entropy as a function of $t$ and
$\epsilon$ for $N=100$.  For fixed $t$, a smaller $\epsilon$ leads to
higher entropy. For $\epsilon=0$ the 2 modules become independent and
the entropy is maximized. 

\medskip

\textbf{Application to a real-world data set.~--}
What about Pearson matrices with entropy in between the maximum and
minimum entropies?  This scenario is, of course, the case of crucial
interest for analyzing empirical data.  As a practical demonstration of
the method's power and its potential applicability, we calculate the
entropy of Pearson correlation matrices obtained from fMRI time courses
of the human brain.  Specifically, we assess the hypothesis that brain
signals display an increase of entropy while under the influence of a
psychedelic~\cite{car14}. For this purpose, we chose the indigenous
beverage ayahuasca~\cite{rib03}, a decoction with rapid antidepressant
effects partially mediated by the serotonergic agonist
\textit{N},\textit{N}-dimethyltryptamine (DMT)~\cite{pal19}.

\begin{figure*}[h!t]
\centering
\includegraphics[scale=0.95]{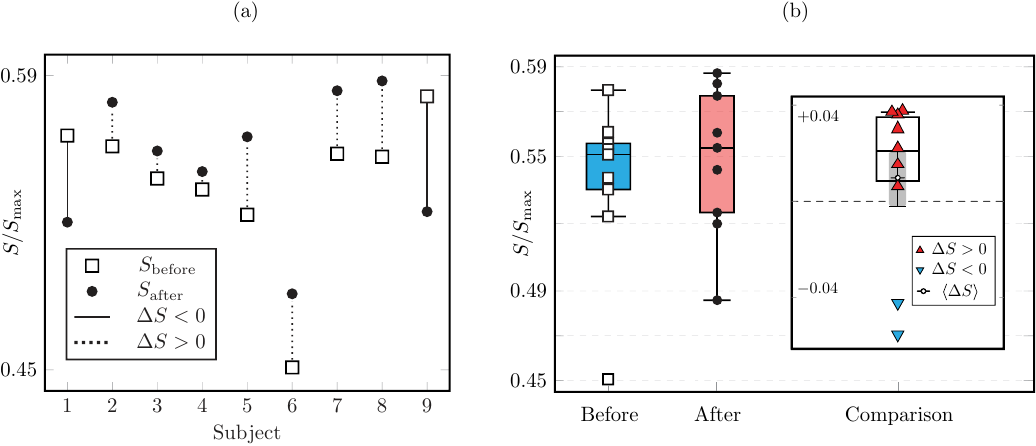}
\caption{
	Entropy before and after ayahuasca.
    (a) 
    The plot shows the entropy $S$ of the nine subjects (horizontal
    axis), both before ($S_{\textrm{before}}$) and after
    ($S_{\textrm{after}}$) ayahuasca ingestion, normalized by
    $S_{\textrm{max}}=\log\left(104\right)$. Entropy variation $\Delta
    S= S_{\textrm{after}}-S_{\textrm{before}}$ was found positive for
    seven subjects (dotted lines).
    (b) 
    Box plots of the distribution $S_{\textrm{before}}$ and
    $S_{\textrm{after}}$, with the single outlier being Subject 6 in the
    condition prior to ayahuasca consumption.  On the right, the inset
    shows the box plot of the distribution of entropy variation. The
    error bar of the average $\langle \Delta S\rangle$ taken over all
    nine subjects is represented by the shaded region, and only slightly
    touches the $\Delta S < 0$ zone.  These findings are broadly
    consistent with the hypothesis of psychedelic-induced brain entropy
    increase, although one must be careful due to the small number of
    subjects involved.
} \label{fig:delta}
\end{figure*}

A total of nine healthy human volunteers were assigned to a neuroimaging
session both before and after ayahuasca intake. We obtained a set of
$N=104$ fMRI time series of the blood-oxygen-level-dependent (BOLD)
signal of the human brain from each subject at both conditions. We next
calculated the density operators. The 18 correlation matrices were
individually scaled according to Eq.~\ref{eq:eq-R-q03475983475} using
$N=104$.  For a detailed description of the acquisition, preprocessing,
and availability of the data, please refer to the 
\hyperlink{supp2}{Supplementary Material}.

In Fig.~\ref{fig:delta} we show the normalized entropy of all 18
matrices $\bm{\rho}$, grouped in terms of subjects (from 1 to 9) for
both experimental conditions (before and after ayahuasca ingestion).
With the exception of two individuals (Subjects 1 and 9), we find
positive entropy variation $\Delta S\equiv
S_{\textrm{after}}-S_{\textrm{before}}$ among subjects (dotted lines).
Although the low number of individuals demands careful analysis in
regards to statistical significance, our findings are in agreement
with previous measures of increased brain entropy due to a psychedelic
ingestion~\cite{tag14,vio17,vio19,leb16,schartner2017}.  Ours,
however, crucially avoids the need of thresholding schemes.  Finally,
measures of brain entropy might differ in the literature, for
instance, by probing either static or dynamical properties of network
links~\cite{keshmiri,drummond}. Here, the entropy proposed is a
measure of the network connectivity repertoire, thus an entropy of the
links.

\medskip

In summary, we have proposed an entirely objective procedure for
directly estimating a unique entropy of a general Pearson matrix
without needing to use thresholding.  We hope that our results will
contribute positively to satisfy, at least partially, the growing and
urgent demand in diverse fields of science for a reliable method of
estimating the information entropy associated with correlations. 

\acknowledgments
A.V.~thanks the support of Progetti di Ricerca di Interesse Nazionale
(PRIN) grant 20174TPEFJ ``TRIPS''. This work was supported by the
Brazilian agencies Conselho Nacional de Desenvolvimento Cient\'{i}fico
e Tecnol\'{o}gico~(CNPq) (Grants No.~304532/2019-3, No.~305062/2017-4,
and No.~302051/2018-0) Coordena\c{c}\~{a}o de Aperfei\c{c}oamento de
Pessoal de N\'{i}vel Superior (CAPES), and Funda\c{c}\~{a}o de Amparo
\`{a} Ci\^{e}ncia e Tecnologia do Estado de Pernambuco (FACEPE).


%
%


\onecolumngrid

\bigskip\medskip

\begin{center}
  \textbf{
      \large Supplementary Material 
}\\[.2cm]
\end{center}

\setcounter{equation}{0}
\setcounter{figure}{0}
\setcounter{table}{0}
\setcounter{page}{1}
\renewcommand{\theequation}{S\arabic{equation}}
\renewcommand{\thefigure}{S\arabic{figure}}
\renewcommand{\thetable}{S\arabic{table}}
\renewcommand{\thepage}{S\arabic{page}}
\renewcommand{\bibnumfmt}[1]{[#1]}
\renewcommand{\citenumfont}[1]{#1}

\hypertarget{supp}{
\subsection*{Guttman Scaling}
}

\subsubsection*{
    The entropy for the homogeneous Guttman-scalable Pearson matrix
}

Whenever a new method of data analysis is developed, it is always
useful to obtain exact analytical results from it, thus aiming to gain
proper insights on the characteristics of the approach.  An idealized,
but useful, type of correlated data is that of a perfect Guttman scale
\cite{guttman-1941,guttman-1950}.  It has been used to study data in
many different areas of knowledge. But in particular, it finds
important usages in health and biomedical related problems
\cite{davies1971, tractenberg2012, argyle2013, maggino2014,
warrian2009}.  We thus shall address a set of correlated data
following the Guttman scaling.

In the homogeneous case~\cite{zwick-1987} (see below), many exact
results are known for the corresponding Pearson correlation matrix
$\bm{R}_{hG}$ \cite{zwick-1987, stober-2015}. As we will show, this
allows to write an analytical expression for the entropy $S_{hG}$ of
$\bm{R}_{hG}$. Although real data rarely fits in with a perfect
Guttman scale, the present $S_{hG}$ could play the role of a null
hypothesis model \cite{zwick-1987}. Indeed, it may be useful for
testing hierarchies as well as features of correlations in distinct
processes involving $N$ populations \cite{kendall-1990, piao-2016}; in
our case all with the same size $n$ and described by ($i=1, \ldots,
N$)
\begin{equation}
  {\boldsymbol x}_i = (x_i^{(1)}, \ldots, x_i^{(n)}) ~.
\label{eq:dataformat}
\end{equation}

\noindent Moreover, Guttman scaling is a very instructive example of
how the entropy portrays data with strong but not absolute
correlations.

For our purposes, we concentrate on dichotomous or binary {\em events}
(or ``items'', a more common terminology in the area of statistical
measurement theory \cite{ zwick-1987,goodman-1975,hand-1996}). Thus,
to each properly defined event $i$ we can associate two distinct
numerical values, 0 and 1. For a given instance or realization pattern
for the collection of events, the numerical value 1 (0) for the event
$i$ means it has (has not) taken place in that pattern. 

The idea underlying the Guttman scaling, also known as cumulative
scaling or scalogram analysis, is to be able to create a table or
indicator matrix --- which should not be confused with the correlation
matrix $\bm{ R}$ itself --- assembling the data in a very particular
manner. If such construction is possible for the data set, the
indicator matrix would display a triangular structure, with all
entries in the lower (upper) part equal to 1 (0), see Table
\ref{tab:table1}.  Hence, within a given pattern of outcomes, there is
a priority of events. If the event $i$ has occurred, then event $i-1$
has also occurred. The Table \ref{tab:table1} explains in more detail
the data organization.

Two extra assumptions for data already observing the Guttman
arrangement are known as homogeneity conditions \cite{zwick-1987}.
They can be stated as the following.  (a) There is no full correlation
between any two populations ${\boldsymbol x}_{i''}$ and ${\boldsymbol
x}_{i'}$.  Actually, in many contexts this is not a really restrictive
hypothesis, either because such coincidence is rare or, if it is not,
for $N$ large enough we can just eliminate the akin ${\boldsymbol
x}_i$ from the data set without loosing relevant statistical
significance.  (b) The probability of obtaining the different
patterns of outcomes indicated in Table \ref{tab:table1} are the
same, \textit{i.e.}, the ${\mathcal N}_i$ in Table
\ref{tab:table1} are essentially all equal.  Arguably, this may
seem a too restricting imposition \cite{stober-2015}.  But perhaps
a bit surprising, it has been verified numerically that certain
results (like principal component analysis and certain properties
of covariance matrices) for homogeneous data do not depart very
much from the case where the ${\mathcal N}_i$ display a normal
distribution (provided the populations have ample variability
\cite{zwick-1987,stober-2015}).

\begin{table*}[b!]
\renewcommand\tablename{Table}
\centering
\caption{\label{tab:table1}
   The data table (or indicator matrix) organized in a specific
   ordering of events $i$, with $i=1,\ldots,N$.  Along each row ---
   representing a particular configuration of events occurrence: a
   pattern --- the entry 1 (0) means that the corresponding event has
   (has not) taken place.  The depicted structure illustrates a
   perfect Guttman scaling: for the pattern $i$, only the events up to
   $i-1$ do take place.  This ascribes a ranking or degree of
   prevalence among the distinct events within a specific pattern.
   ${\mathcal N}_i$ is the number of times the pattern $i$ is found in
   the full data set collection. \\
}
\begin{tabular}{ccccccccc}
\renewcommand\tablename{Table}
 Pattern & Event 1 & Event 2 & Event 3 & Event 4 & Event 5 &
  \ldots & Event $N$
  & Number of occurrences \\
\hline
1      & 0 & 0 & 0 & 0 & 0 & \ldots & 0 & ${\mathcal N}_1$ \\
2      & 1 & 0 & 0 & 0 & 0 & \ldots & 0 & ${\mathcal N}_2$ \\
3      & 1 & 1 & 0 & 0 & 0 & \ldots & 0 & ${\mathcal N}_3$ \\
4      & 1 & 1 & 1 & 0 & 0 & \ldots & 0 & ${\mathcal N}_4$ \\
5      & 1 & 1 & 1 & 1 & 0 & \ldots & 0 & ${\mathcal N}_5$ \\
6      & 1 & 1 & 1 & 1 & 1 & \ldots & 0 & ${\mathcal N}_6$ \\
\vdots & \vdots & \vdots & \vdots & \vdots & \vdots
& \vdots &  0 & \vdots \\
$N+1$  & 1 & 1 & 1 & 1 & 1 & \ldots & 1 & ${\mathcal N}_{N+1}$\\
\hline
\end{tabular}
\end{table*}

Therefore, it is correct to affirm that data obeying the homogeneous
Guttman scaling is not ubiquitous. But the relevant fact for our
purposes is that, precisely because of that, the Pearson correlation
matrix $\bm{R}_{hG}$ can be derived analytically \cite{zwick-1987}.
Furthermore, its exact eigenvalues $\tilde{\lambda}_i$ have been
obtained in \cite{zwick-1987}.  Since the eigenvalues ${\lambda}_i$ of
$\bm{\rho}_{hG}$ are simply $\tilde{\lambda}_i/N$, then from
\cite{zwick-1987} we get
\begin{equation}
  \lambda_i = \frac{1+N^{-1}}{i \, (i+1)}~.
\end{equation}

\begin{figure}[ht!]
\setcounter{figure}{0}
\renewcommand\figurename{Fig.}
\centering
\includegraphics[width=0.75\textwidth]{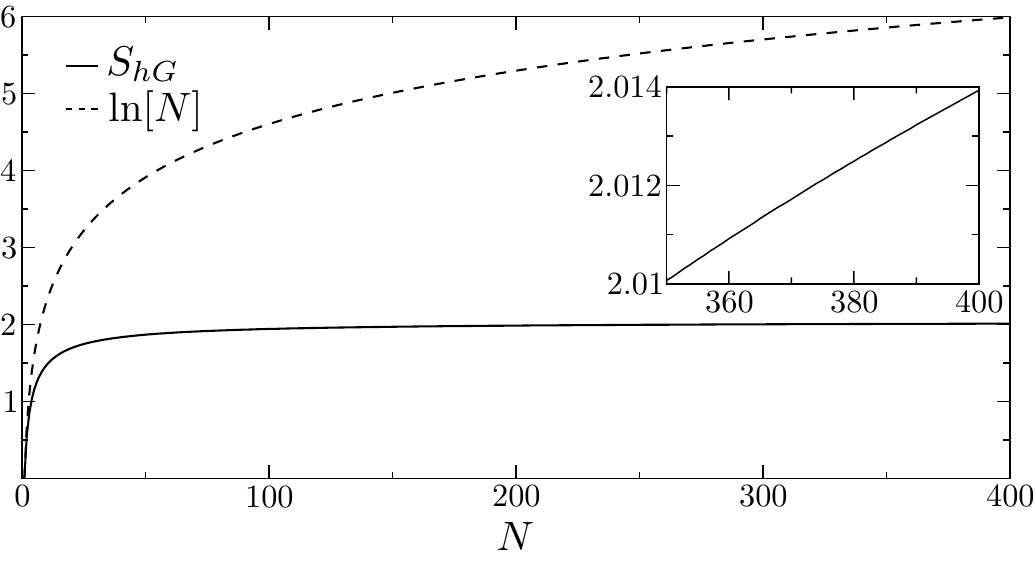}
\caption{
	The entropy	for a Pearson correlation matrix resulting from an 
	ideal homogeneous Guttman-scalable data set.
    The entropy $S_{hG}$ tends to a constant value for the population
    (number of items) $N$ going to infinity (see the inset). For
    comparison, it also shows the entropy for a completely uncorrelated
    data set, for which $S=\ln[N]$.
} 
\label{fig1}
\end{figure}

Notice that the $\lambda_i$ depend only on a single parameter, the
population number $N$.  So, finally we find
\begin{align}
  S_{hG} &= - (1+N^{-1}) \sum_{i=1}^N \, \frac{1}{i \, (i+1)}
  \ln\left[\frac{1+N^{-1}}{i \, (i+1)} \right]
\nonumber
  \\
&= \nonumber
 \ln[N] - (1-N^{-2}) \ln[N+1]
\\
& \quad
+ 2 \, (1+N^{-1})
  \sum_{i=1}^{N-1} \frac{1}{i \, (i+2)} \ln[i + 1]~.
\end{align}
Since it is well known that the series $\sum_{i=1}^{\infty} \ln[i]/i^2$
converges, it is not difficult to prove that $S_{hG}$ also converges as
$N \rightarrow \infty$.  Also, due to the peculiar way items are
correlated in a homogeneous Guttman-scalable data collection, after a
rapid increase with $N$, the entropy $S_{hG}$ tends to a plateau as $N$
increases.  Figure~\ref{fig1} illustrates the general behavior of
$S_{hG}(N)$, also comparing it with the entropy for a Pearson
correlation matrix of a totally uncorrelated data set, for which
$S=\ln[N]$.

\subsection{Possible ways to frame a general data
set in terms of ``events''}

An important question is certainly how a given data set, given as in
Eq.~(\ref{eq:dataformat}), can be described in terms of events and thus
to allow an indicator matrix representation.  Below we summarize two
possible procedures to do so.

The first is when we can associate phenomenological behavior to the
data.  Then, we can define an event by classifying {\em appropriate}
(\textit{i.e.}, easier to characterize in terms of ranking of
predominance) information which can be inferred once we know the
explicit values of the $x_i^{(j)}$.  For example, suppose whenever
$x_i^{(j)}$ is above a threshold $\gamma$, we can conclude it has been a
high electrical pulse crossing the population $i$ at the measurement
step $j$.  Then, an event could be the existence (1) or not (0) of an
intense pulse.  Further, suppose that a determined biochemical reaction
takes place when $\alpha < x_i^{(j)} < \beta$.  Such reaction could also
constitute another event, with the values 0 and 1 being ascribed
conforming to the numerical intervals for the $x_i^{(j)}$.  Importantly,
if $\langle {\boldsymbol x}_i \rangle < \gamma < \alpha$, in principle
we may expect an order of priority between the high pulse and the
biochemical reaction.  In this way, we might try to organize the
events as in Table \ref{tab:table1} according to the numerical
sequences in Eq.~(\ref{eq:dataformat}).  Of course, such an approach
is heavily based on the concrete understanding of the physical,
chemical, biological, etc., aspects underlying the problem studied.
In the absence of this type of knowledge, the protocol is not
feasible.

A second possibility is to consider an algorithmically oriented method,
not relying on any {\em a priori} qualitative information about the
process.  But then, admittedly it has a higher chance of not satisfying
the perfect homogeneous Guttman scaling \cite{stober-2015,coombs-1978}.
For the full data set, let us define $a$ and $b$ as the minimum and
maximum values assumed by the entire collection of $x_i^{(j)}$.  Thus,
we divide $[a,b]$ into $N+1$ intervals $\Delta_0, \Delta_1, \ldots ,
\Delta_N$.  Here, $\Delta_0$ must represent the most frequent numerical
interval for the data set, with the other $\Delta_i$ presenting a
decreasing order of occurrence with $i$.  The events to be considered in
the indicator matrix are the occurrence of the intervals $\Delta_1,
\Delta_2, \ldots, \Delta_N$ ($\Delta_0$ is not explicitly included in
the events table, but it contributes with the ``0'''s in it).  For $M_i$
the number of times the interval $\Delta_i$ appears in the data set, we
have $\sum_{i=0}^N M_i = n \times N$.  All the incidences $M_i$
($i=0,1,\ldots,N$) of the events $\Delta_i$ should thus be combined to
generate a structure like that in Table \ref{tab:table1}.  An
illustration of how to proceed in the particular case the data allows an
exact homogeneous Guttman scaling is depicted in Table \ref{tab:table2}.
Finally, given that there is a certain freedom in choosing the intervals
$\Delta_i$, this must be done in such way to get an indicator matrix as
close as possible to the pattern in Table \ref{tab:table1}.

\begin{table*}[t!]
\renewcommand\tablename{Table}
\centering
  \caption{\label{tab:table2}
    Consider that for a data set of $N=3$ and $n=4$, the twelve
    $x_i^{(j)}$ assume $M_0 = 6$, $M_1 = 3$, $M_2 = 2$, $M_3 = 1$
    values, respectively, in the numerical intervals $\Delta_0$,
    $\Delta_1$, $\Delta_2$ and $\Delta_3$.  Then, taking as the events
    the instances (irrespective of $i$ and $j$) in which the variables
    are within $\Delta_1$, $\Delta_2$ and $\Delta_3$, one can construct
    an indicator matrix exactly satisfying a homogeneous Guttman
    scaling. Indeed, the three, two and one ``1'''s in the columns Event
    1, Event 2 and Event 3, correspond to $M_1$, $M_2$ and $M_3$,
    whereas the total of six ``0'''s spread in the whole table are due
    to $M_0$. \\
}
\begin{tabular}{ccccc}
  Pattern & Event 1: Interval $\Delta_1$ & Event 2: Interval $\Delta_2$ &
  Event 3: Interval $\Delta_3$ &
  Occurrence of the pattern \\
\hline
1      & 0 & 0 & 0 & 1 \\
2      & 1 & 0 & 0 & 1 \\
3      & 1 & 1 & 0 & 1 \\
4      & 1 & 1 & 1 & 1 \\
\hline
\end{tabular}
\end{table*}

\hypertarget{supp2}{
\section*{Methods}
}

\subsection*{Data acquisition}

A total of nine healthy adults (five women) with no history of
neurological or psychiatric disorders (assessed by DSM-IV structured
interview), volunteered to ingest 120–200\,mL (2.2\,mL/kg of body
weight) of ayahuasca known to contain 0.8\,mg/mL of its main
psychoactive compound DMT and 0.21\,mg/mL of harmine, a
$\beta$-carboline which allows for the entrance of DMT into the
bloodstream \cite{dra12}. The volunteers were assigned with two
task-free fMRI sessions: the first one prior to the ayahuasca
ingestion, and the second one at 40 minutes after the brew intake
(when ayahuasca's effects begin to occur, lasting for approximately
four hours). Both sessions required the participants to maintain an
awake resting state. The experimental procedure was approved by the
Ethics and Research Committee of the University of São Paulo at
Ribeirão Preto (No.~14672/2006).  Written informed consent was
obtained from all volunteers. All experimental procedures were
performed in accordance with the relevant guidelines and regulations.

The fMRI images were acquired in a 1.5\,T scanner (Siemens, Magneton
Vision), using an EPI-BOLD sequence to obtain 150 volumes with
parameters TR = 1700\,ms, TE = 66\,ms, FOV = 220\,mm, matrix $64\times
64$, and voxel dimensions of
1.72\,mm\,$\times$\,1.72\,mm\,$\times$\,1.72\,mm. The images were
preprocessed in the FSL software \cite{jen12}, and consisted of
slice-timing, head motion, and spatial smoothing corrections (Gaussian
kernel, FWHM = 5\,mm). Nine regressors were used within a general
linear model (GLM): six regressors to movement correction; one to
white matter signal; one to cerebrospinal fluid; and one to global
signal. The images were normalized to the standard anatomical space of
the Montreal Neurological Institute (MNI152 template)~\cite{bret02}. 

The preprocessed fMRI images were then parcellated into 110 anatomical
regions of interest (ROIs) in accordance with the Harvard-Oxford
cortical and subcortical atlas (see Supplementary Table S3).  Due to
acquisition limitations, six regions were excluded from further
analysis. Thus, for each one of the remaining 104 ROIs, we averaged
the BOLD signal of the regions' associated voxels, resulting in time
courses listing a sequence of $n=150$ data points. To reduce
confounders in the signal, a maximum overlap discrete wavelet
transform (MODWT) was applied to the series in order to select the
typical frequency range (0.01–0.1\,Hz) of the resting state signal
\cite{oni14}. Finally, in possession of the $N=104$ time series
$\bm{x}_i$, we pairwise-correlated them using the Pearson correlation
coefficient of equation~\ref{eq:rij}, resulting in the $104\times 104$
Pearson correlation matrix $\bm{R}$. We thus ended up with a total of
18 functional connectivity networks: two conditions (before and after
ayahuasca) for each of the nine subjects involved.

\subsection*{Subjective evaluation}

Individual responses of the Clinician-Administered Dissociative States
Scale (CADSS), a psychometric scale that can discriminate subjects
with dissociative disorders \cite{bremner}, were collected and their
scores were plotted against the respective entropy. All subjects
showed an increase in depersonalization as assessed by the CADSS scale
(see Supplementary Figure S2).

\section*{Data availability}
The datasets generated and/or analysed during the current study are
available in the GitHub repository,
\linebreak\href{https://github.com/hfelippe/arXiv.2106.05379}
{https://github.com/hfelippe/arXiv.2106.05379}.  In-house scripts to
generate the correlation matrices and figures are also available in
the referred repository.  The original fMRI data can be sent upon
request to DBA
(\href{mailto:draulio@neuro.ufrn.br}{draulio@neuro.ufrn.br}).

\clearpage

\begin{figure}
\includegraphics[width=0.75\textwidth]{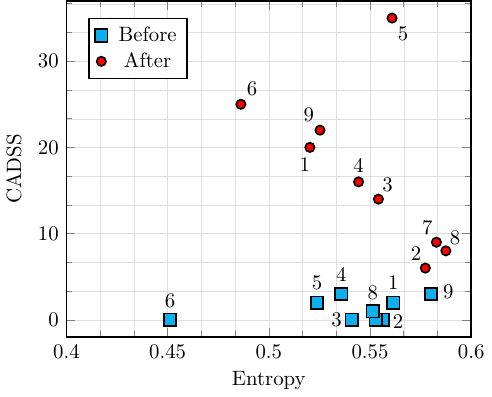}
\caption{
	Entropy against CADSS scores.
    Blue squares refer to the condition prior to the ayahuasca
    ingestion, whereas red circles is posterior to the consumption of
    ayahuasca. The numbers 1 to 9 index the nine Subjects,
    respectively. We note that Subjects 1 and 9 were the only
    individuals who showed a decrease in entropy after ingestion of
    ayahuasca (a translation to the left on the graph).  However, all
    subjects scored higher in the Clinician-Administered Dissociative
    States Scale (CADSS).
} 
\label{figs2}
\end{figure}

\clearpage

\begin{table*}[h!t]
\caption{
    Brain regions in accordance with the Harvard-Oxford cortical atlas.  
    Asterisks ($^*$) are assigned to the six excluded regions.
    Ordered pairs $(i, j)$ refers to the left ($i$) and right ($j$)
    hemispheres of the corresponding region. \vspace{0.125cm}
}
\label{tablebrain}

\begin{tabular}{l | c}
         Index & Region \\ \hline
        $(1,2)$ & \; Frontal Pole \\
        $(3,4)$ & \; Insular Cortex \\
        $(5,6)$ & \; Superior Frontal Gyrus \\
        $(7,8)$ & \; Middle Frontal Gyrus \\
        $(9,10)$ & \; Inferior Frontal Gyrus, pars triangularis \\
        $(11,12)$ & \; Inferior Frontal Gyrus, pars opercularis \\
        $(13,14)$ & \; Precentral Gyrus \\
        $(15,16)$ & \; Temporal Pole \\
        $(17,18)$ & \; Superior Temporal Gyrus, anterior division \\
        $(19,20)$ & \; Superior Temporal Gyrus, posterior division \\
        $(21,22)$ & \; Middle Temporal Gyrus, anterior division \\
        $(23,24)$ & \; Middle Temporal Gyrus, posterior division \\
        $(25,26)$ & \; Middle Temporal Gyrus, temporooccipital part \\
        $(27,28)*$ & \; Inferior Temporal Gyrus, anterior division \\
        $(29,30)$ & \; Inferior Temporal Gyrus, posterior division \\
        $(31,32)$ & \; Inferior Temporal Gyrus, temporooccipital part \\
        $(33,34)$ & \; Postcentral Gyrus \\
        $(35,36)*$ & \; Superior Parietal Lobule \\
        $(37,38)$ & \; Supramarginal Gyrus, anterior division \\
        $(39,40)$ & \; Supramarginal Gyrus, posterior division \\
        $(41,42)$ & \; Angular Gyrus \\
        $(43,44)$ & \; Lateral Occipital Cortex, superior division \\
        $(45,46)$ & \; Lateral Occipital Cortex, inferior division \\
        $(47,48)$ & \; Intracalcarine Cortex \\
        $(49,50)$ & \; Frontal Medial Cortex \\
        $(51,52)$ & \; Juxtapositional lobule cortex \\
        $(53,54)$ & \; Subcallosal Cortex \\
        $(55,56)$ & \; Paracingulate Gyrus \\
        $(57,58)$ & \; Cingulate Gyrus, anterior division \\
        $(59,60)$ & \; Cingulate Gyrus, posterior division \\
        $(61,62)$ & \; Precuneous Cortex \\
        $(63,64)$ & \; Cuneal Cortex \\
        $(65,66)$ & \; Frontal Orbital Cortex \\
        $(67,68)$ & \; Parahippocampal Gyrus, anterior division \\
        $(69,70)$ & \; Parahippocampal Gyrus, posterior division \\
        $(71,72)$ & \; Lingual Gyrus \\
        $(73,74)*$ & \; Temporal Fusiform Cortex, anterior division \\
        $(75,76)$ & \; Temporal Fusiform Cortex, posterior division \\
        $(77,78)$ & \; Temporal Occipital Fusiform Cortex \\
        $(79,80)$ & \; Occipital Fusiform Gyrus \\
        $(81,82)$ & \; Frontal Operculum Cortex \\
        $(83,84)$ & \; Central Opercular Cortex \\
        $(85,86)$ & \; Parietal Operculum Cortex \\
        $(87,88)$ & \; Planum Polare \\
        $(89,90)$ & \; Heschl's Gyrus (includes H1 and H2) \\
        $(91,92)$ & \; Planum Temporale \\
        $(93,94)$ & \; Supracalcarine Cortex \\
        $(95,96)$ & \; Occipital Pole \\
        $(97, 104)$\, & \; Thalamus \\
        $(98, 105)$\,  & \; Caudate \\
        $(99, 106)$\, & \; Putamen \\
        $(100, 107)$\, & \; Pallidum \\
        $(101, 108)$\, & \; Hippocampus \\
        $(102, 109)$\, & \; Amygdala \\
        $(103, 110)$\, & \; Accumbens
\end{tabular}

\end{table*}

\clearpage

\end{document}